# Short-range antiferromagnetic interaction and spin-phonon coupling in $La_2CoMnO_6$ double perovskite

*R. X. Silva[1,6*], C. C. Santos[2], H. Reichlova[3], X. Marti[3,4] R. Paniago[5] and C. W. A. Paschoal[6]*

[1]Coordenação de Ciências Naturais, Universidade Federal do Maranhão, Campus VII, 65400-000, Codó-MA, Brazil

[2]Departamento de Física, Universidade Federal do Maranhão, CCET, São Luís, MA 65080-805, Brazil

[3] Institute of Physics ASCR, v.v.i., Cukrovarnická 10, 162 53 Praha 6, Czech Republic

[4] Centre d'Investigacions en Nanociencia I Nanotechnologia (CIN2), CSIC-ICN, Bellaterra 08193, Barcelona, Spain

[5]Departamento de Física, Universidade Federal de Minas Gerais, ICEx, 31270-901 Belo Horizonte-MG, Brazil

[6] Departamento de Física, Universidade Federal do Ceará, Campus do Pici, PO Box 6030, 60455-970, Fortaleza - CE, Brazil.

[*] Corresponding author. Tel: +55 98 982234219

E-mail address: rosivaldo.xs@ufma.br (R. X. Silva)

# *Abstract*

Weak antiferromagnetic (AF) interaction in the ferromagnetic (FM) partially ordered $La_2CoMnO_6$ (LCMO) was detected by Raman spectroscopy by monitoring spin-phonon coupling. Because of the sensibility to probe short-range disorder and lattice modifications, the Raman spectroscopy showed to be an useful tool to indicate less remarkable magnetic transitions in LCMO compound. Apart from the expected spin-phonon coupling due to the long-range FM superexchange (Tc ~230 K), phonons parameters pointed out an additional spin-phonon coupling related to the short-range AF interaction at Tc ~135 K, which was not detected from the magnetic bulk response. These results reinforce the Raman spectroscopy as a powerful technique to detect antisite disorder into $A_2B'B''O_6$ magnetic double perovskites, whose magnetic properties are driven by superexchange interactions.

## 1. Introduction

Multiferroics with a crystalline structure derived from the $RE_2Me^{2+}MnO_6$ family of double perovskites, where RE and $Me^{2+}$ are rare earth and transition metal ions, respectively, have attracted much attention due to their peculiar electric and magnetic properties, which involve giant dielectric constant (>$10^3$) [1,2], magnetoresistance [3], magnetocaloric and magnetodielectric effect [3–5], relaxor behavior [6,7], among others. $La_2CoMnO_6$ (LCMO) and $La_2NiMnO_6$ (LNMO) are good examples of these materials, which, in addition to presenting the above characteristics, have a Curie point near room temperature, around 230K and 280K, respectively, so they are suggested for applications in electrically readable magnetic data storage, spintronic devices and, more recently, for high-performance supercapacitors [3,8–10].

Structural disorder in complex perovskites plays an essential role because it strongly affects their physical and chemical properties. For example, phonons and, consequently, the dielectric response (in the microwave electromagnetic range) are strongly influenced by the antisite disorder in perovskites with 1:2 cation order ($A_3BB'_2X_9$) [11–14]. Also, in $Ba_3CaNb_2O_9$, the high antisite disorder favors the conductivity, enabling its application as protonic conductors [15]. Particularly in $RE_2Me^{2+}MnO_6$ perovskites, the order is critical because, in such materials, the ferromagnetic behavior is driven by the superexchange interaction between $Me^{2+}$ (partially filled $e_g$ orbital) and $Mn^{4+}$ (filled $t_{2g}$ orbital) mediated by the oxygen. Thus, disorder generates antisite defects of the type $Me^{2+}$ – O – $Me^{2+}$ and $Mn^{4+}$ – O – $Mn^{4+}$, which induce short-range antiferromagnetic interactions, decreasing the overall ferromagnetic state. High ordered $RE_2Me^{2+}MnO_6$ single crystals exhibit narrow hysteresis (or low coercive field) and only one ferromagnetic (FM) transition, showing only a smooth magnetization at low temperatures.

Commonly, the coexistence of two phases, oxygen vacancies, mixed-valence and antisite defects promotes multiple magnetic interactions at low temperatures, as well as a low saturation magnetization and high coercive field. In the case of LCMO, one of the most ordered polycrystalline sample reported had a magnetization of 5.89 $\mu_B$/f.u. (formula unit), really close to the theoretical saturation magnetization of 6.0 $\mu_B$/f.u for a fully ordered sample, a relatively weak coercive field of 2.9 kOe, and only one magnetic transition at $T_c$ = 235 K[16]. On the other hand, LCMO samples processed at very high temperatures (T>1200 °C) tend to be oxygen-deficient [17–22]. The presence of oxygen vacancies introduces $Mn^{3+}$ and $Co^{3+}$ íons which interact ferromagneticly through a vibronic $e^1$-O-$e^1$ superexchange interaction with small stabilization and lower Curie temperature ($T_{c2}$) and high coercitivity (~10 kOe) [16,18,23–25]. Moreover, structural ordering influences dielectric [26,27], vibrational [16,19] and electronic properties [21]. Wherein, oxygen content, charges, and antisite defects can be tuned by the synthesis conditions [16–19,26,27].

Raman scattering is a powerful probe to detect the disorder in 1:2 complex perovskites because the two-phonon mode behavior is usually observed, which is induced by the disorder and characterizes the presence of the antisite defects. However, the observation of antisite occupation in 1:1 ordered samples, mainly in $RE_2Me^{2+}MnO_6$ perovskites, is difficult because of the absence of the two-phonon mode behavior. In fact, due to the charge difference between $Me^{2+}$ and $Mn^{4+}$ ions, the octahedral modes are related mainly to the $Mn^{4+}$ ions, as reported by Silva *et al.* [28]. The two-mode behavior is only observed when there is a high difference in mass of the metallic ions [29], which is not the case for $RE_2Me^{2+}MnO_6$ perovskites. For LCMO, for example, the spectra of the ordered and disordered samples are very similar, as can be verified in Ref. [16]. However, these compounds usually present strong coupling between magnetic order and phonons, which is easily probed by the Raman spectroscopy through both the position and linewidth of the phonon associated with the $MnO_6$ stretching [30–36].

In this paper, we exploit the coupling between phonons and magnetic ordering to detect low concentration of antisite defects in magnetic double perovskites using Raman spectroscopy. Thus, this analysis proved to be efficient in detecting the competition between the main FM ordering and short-range antiferromagntic interactions. For this purpose, we used a partially disordered LCMO ceramic obtained by the polymeric precursor method as a prototype sample. To support our proposal, magnetic, X-ray Photoelectron Spectroscopy (XPS) and X-ray diffraction techniques were also employed.

## 2. Experimental

Polycrystalline samples of LCMO were produced through the polymeric precursor method [16,37][37]. Wherein cobalt acetate tetrahydrate ($C_4H_6CoO_4 \cdot 4H_2O$, Sigma Aldrich), manganese nitrate hydrate ($MnN_2O_6 \cdot xH_2O$, Sigma Aldrich) and lanthanum oxide ($La_2O_3$, Sigma, Aldrich) were added to an aqueous solution in stoichiometric amounts. The mixture was kept under stirring for about 2 h to obtain a resin, followed by heating treatment at 400 °C to remove the solvent and the organic matter. The starting powder was ground in an agate mortar and heated at 1100 °C for 16 h. Then, the obtained powder was reground and annealed for 16 h at 1100 °C to obtain the LCMO sample. The crystalline structure of the sample was analyzed by applying the X-ray powder diffraction technique using a Bruker D8 Advance with Cu-Kα radiation (40 kV, 40 mA) over the range from 20° to 100° (0.02 °/step with 0.3 s/step).The structure was refined using the GSAS code [38–40][39,40]. X-ray Photoelectron Spectroscopy (XPS) measurements were performed in a VG ESCALAB 220i-XL system, using Al-Ka radiation and under the pressure of $2\times10^{-10}$ mbar. Survey XPS spectra were collected with a pass energy of 50 eV, and detailed spectra around the Co 2p and Mn 2p regions were taken with 20 eV pass energy. Magnetic measurements were carried out using a Superconducting Quantum Interference Device (SQUID - Quantum Design), and the temperature sweeps were collected with 4 cm long Reciprocating Sample Option scans. Raman spectroscopy measurements were performed using a Jobin-Yvon T64000 Triple Spectrometer coupled to an Olympus Microscope (model BX-41) equipped with a 20x (long work distance) achromatic lens, which was configured in backscattering geometry. The 514.5 nm line of an Innova Coherent laser operating at 18 mW was used to excite the Raman signal, which was collected in a $N_2$-cooled CCD detector. All slits were set up to attain a resolution lower than 1 $cm^{-1}$. Low-temperature measurements were performed by using a closed-cycle He cryostat with a temperature control accuracy of 0.1 K.

## 3. Results and Discussion

The X-ray powder diffraction pattern obtained from LCMO sample at room temperature is shown in Fig. 1 (a). It was indexed based on a monoclinic unit cell with symmetry belonging to the space group $P2_1/n$, coherent with a quite ordered sample [16,18]. The refined structure parameters are in excellent agreement with both previous studies in isostructural compounds obtained by solid-state reaction [41] and theoretical results calculated by the SPuDS program [42].

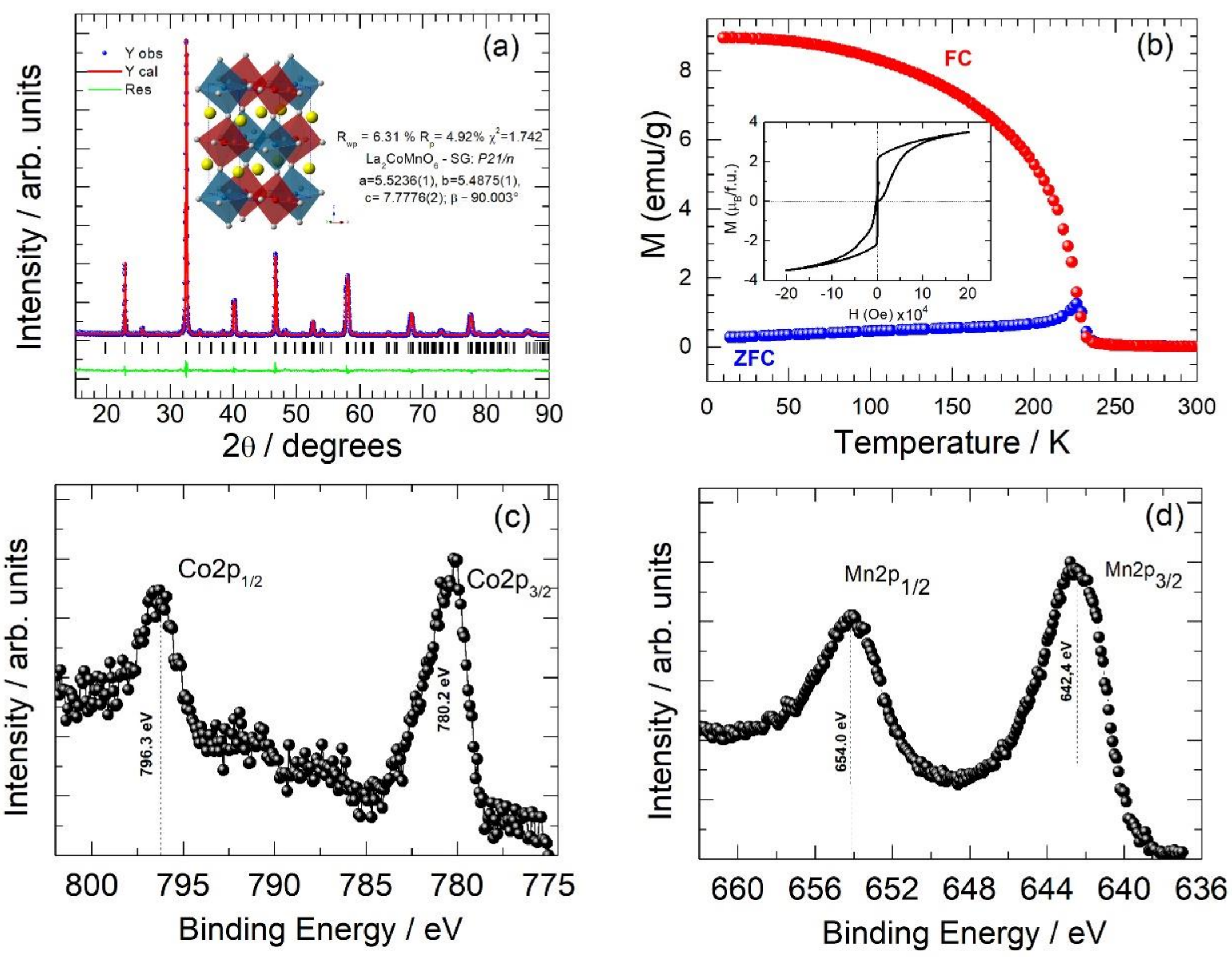

**Fig. 1** – (a) X-ray powder diffraction pattern of LCMO ceramic. The solid red line is the fit using the Rietveld method, and the solid green line is the residual between the experimental and calculated patterns. (b) FC and ZFC magnetization with a field of 100 Oe. The inset shows the hysteresis curve at 10 K. (c) XPS spectra of the partially disordered LCMO (black filled squares) for the (c) Co 2p and (d) Mn 2p energy regions.

Fig. 1 (b) shows the temperature-dependent magnetization of LCMO. The Field-cooled (FC) and the Zero-field-cooled (ZFC) magnetization data were acquired under a field of 100 Oe (10 mT). These magnetization curves are irreversible below T < 226 K, and the apparent divergence between FC and ZFC curves is an indication of the magnetic frustration due to competition between the ferromagnetic and antiferromagnetic interactions, implying in a spin-glass behavior [43]. The magnetization curve reveals the onset of the net magnetic moment near 226 K, i.e., the Curie temperature, consistent with previous magnetic measurements in LCMO [16,44,45], and any second FM transition was observed [46,47].The inset of Fig. 1(b) shows the hysteresis loop at 10 K, and magnetization value tends towards the saturation magnetization value of ~ 3.49 $\mu_B$/f.u. An abrupt change in $H_C$ (coercive field) is observed. Similar effects have been observed in isostructural and related perovskites, where the inhomogeneous concentration of defects and antiphase boundaries (APB) domain borders lead to the strongly pinned magnetic domain walls [48–50].

Magnetic measurements have been used to estimate the B-site structural order in LCMO. Previous works have shown that the structural order level of B-site depends sensitively on their synthesis conditions and can be adjusted as a function of the calcination temperature in air atmosphere [16,21], wherein the sample obtained at 1000 °C had the best order level, with Tc ~235K and saturation magnetization of 5.89 $\mu_B$/f.u. (matching to a nearly stoichiometric ordered sample that contains $Co^{2+}$ and $Mn^{4+}$ cations) [16,21]. According to Blasse's model [51], [51]which purposes to estimate the structural B site ordering using the experimental and theoretical saturation magnetic moments ratio (called δ), this magnetization would correspond to δ =0.98, or about 1% of antisite defects. Thus, for observed saturation magnetization of 3.49 $\mu_B$/f.u. (see inset of Fig. 1(b)), we have a δ=0.58, corresponding to 21% of antisite defects. Despite the high number of antisite defects, the obtained magnetization curve shows only a prominent ferromagnetic transition at 226 K. This transition is usually attributed to long-range ferromagnetic (FM) interaction due to the superexchange interaction between $Mn^{4+}$-O-$Co^{2+}$, stabilized by a stoichiometric oxygen concentration. Dass *et al.*[18] showed that the stoichiometric compound has a single evident magnetic phase with $T_c$ = 226 K (high $T_c$) and $M_s$= 5.78(2) $\mu_B$/f.u. Oxygen deficiencies stimulate the appearance of $Co^{3+}$ ($t^5e^1$) and Jahn-Teller $Mn^{3+}$ ($t^3e^1$) ions, which generate a vibrionic superexchange FM interaction, but with a lower exchange stabilization and a lower critical temperature (low $T_c$). Multiple magnetic transitions are common in compounds heated above 1200°C in air [17–22].

XPS measurements were carried out to evaliate the oxidation states of Co and Mn ions in partially ordered LCMO as observed in Fig. 1 (c) and (d). Fig. 1(c) shows the Co 2p spectrum with the prominent peaks located at 780.2 eV, for Co $2p_{3/2}$, and 796,3 eV, for Co $2p_{1/2}$, whereas the Co $2p_{3/2}$ peak is observed at 780.5 eV for CoO [52] and the weaker satellite peak observed at 787.1 eV are also features of CoO [53]. Thus, this analysis indicates the dominance of $Co^{2+}$ species [54,55]. Fig. 1(d) shows the Mn 2p spectrum. Considering that the main XPS peak of $MnO_2$, Mn $2p_{3/2}$ peak, is observed at 642.2 eV [52,55,56], the peaks observed at about 642.4 eV and 654.0 eV can be associated with Mn $2p_{3/2}$ and Mn $2p_{1/2}$, respectively. Also, these result it is very similar that reported for a very high ordered polycrystalline LCMO sample [16], and together with the magnetic measurement analysis, ruled out the possibility of a significant amount of $Co^{3+}$ and $Mn^{3+}$. Moreover, they confirm that the low saturation magnetization is due to the antiferromagnetic interactions (AFM) promoted by defects of antisite $Mn^{4+}$−O−$Mn^{4+}$ or $Co^{2+}$−O−$Co^{2+}$, and the interaction between APB that occur between regions with AFM interactions, $Mn^{4+}$−O−$Mn^{4+}$ or $Co^{2+}$−O−$Co^{2}$, or between FM regions promoting AFM interactions in the contours.

The room-temperature Raman spectrum of the LCMO is shown in Fig. 2(a). The spectrum is typical of a manganite with a double perovskite structure and is in excellent agreement with previous works [19,57]. According to Ref. [57], the more intense phonon band, at 646 $cm^{-1}$, corresponds to the ($BO_6$) octahedral symmetric stretching (SS) mode, whereas the mode at 499 $cm^{-1}$ is assigned as a combination of both asymmetric stretching (AS) and bending modes. The low-intensity vibrational modes, observed at 426 and 254 $cm^{-1}$, are associated with the octahedral bending mode, with La moving in the xz plane, and the octahedra out-of-phase tilting vibration along the x-axis, with La and apical O moving in the xz plane, respectively. Furthermore, the two modes at high wavenumbers, i.e., at 1160 and 1283 $cm^{-1}$, are related to the combination of the symmetrical stretching and bending modes and the symmetrical stretching vibration overtone, respectively. The low-intensity modes are shown in the inset of Fig. 2(a).

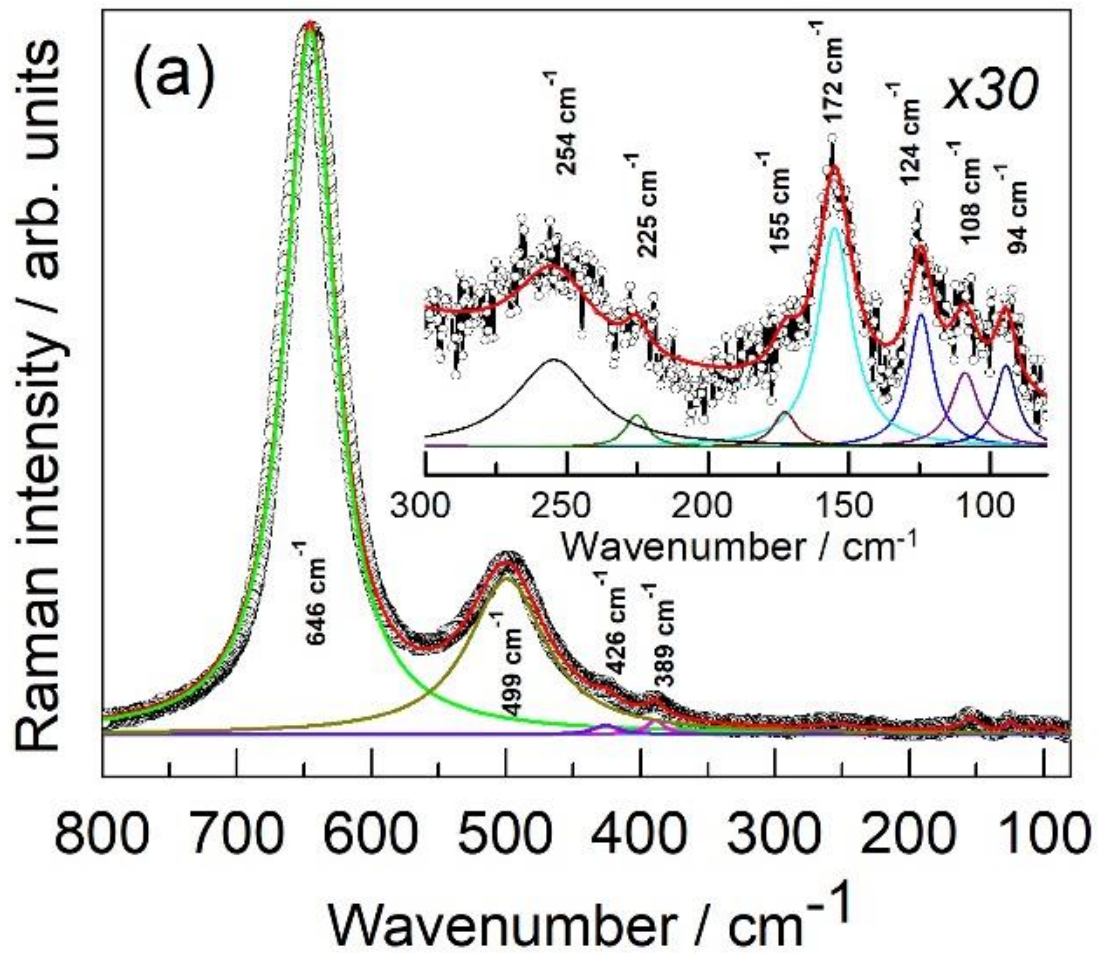

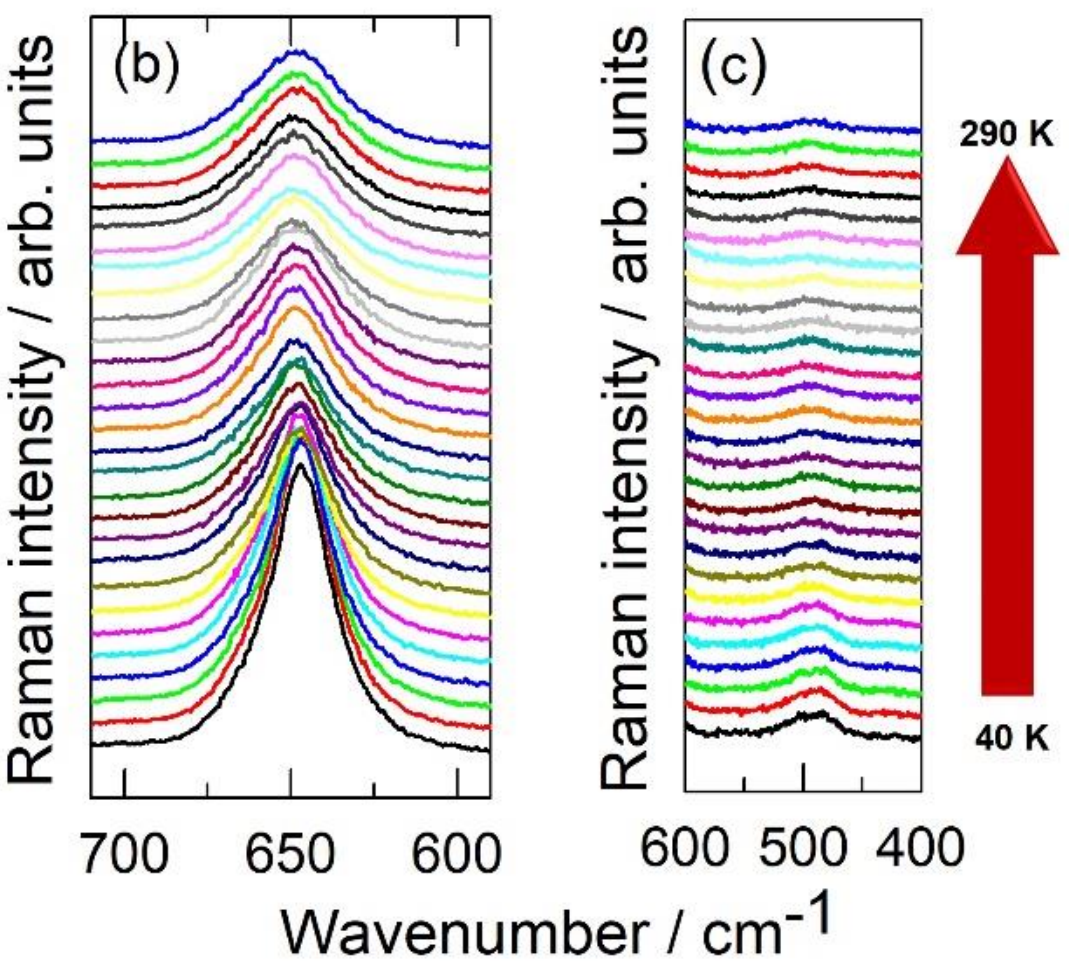

**Fig. 2** – Raman spectrum of LCMO at room temperature. The inset shows the low wavenumber region of the spectrum. Temperature-dependent Raman-active phonon spectra of LCMO in the wavenumber range of (b) (Co/Mn)$O_6$ symmetric stretching (SS) region and (c) (Co/Mn)$O_6$ asymmetric stretching (AS) and bending mode.

The temperature-dependent Raman spectra of LCMO between room temperature and 40 K are shown in Figs. 2 (b-c). Usually, a spin-phonon coupling in similar double perovskites is observed in octahedral stretching vibrational modes [30,32–34]. The temperature dependence of these modes is shown in Fig. 3 (a-b). The expected anharmonic contribution to the temperature dependence of a phonon position modeled by Balkanski [58] is given by

$$\omega(T) = \omega_o - C\left[1 + \frac{2}{(e^{\hbar\omega_o/k_B T} - 1)}\right]$$

with $C$ and $\omega_o$ being the fitting parameters. In the absence of structural phase transitions, as it happens in LCMO, the temperature dependence of the phonons must follow this behavior. However, it is clear that the stretching phonon exhibits a marked deviation from the anharmonic model, which is similar to that recently observed in LCMO [59], as well as for other double perovskites with magnetic transitions, and it is associated with the renormalization of the phonons induced by the magnetic ordering [30,32–35,57,60,61]. The coupling temperature coincides with the start of the magnetic ordering, as shown by the spontaneous magnetization of LCMO ( see Fig. 1 (b)).

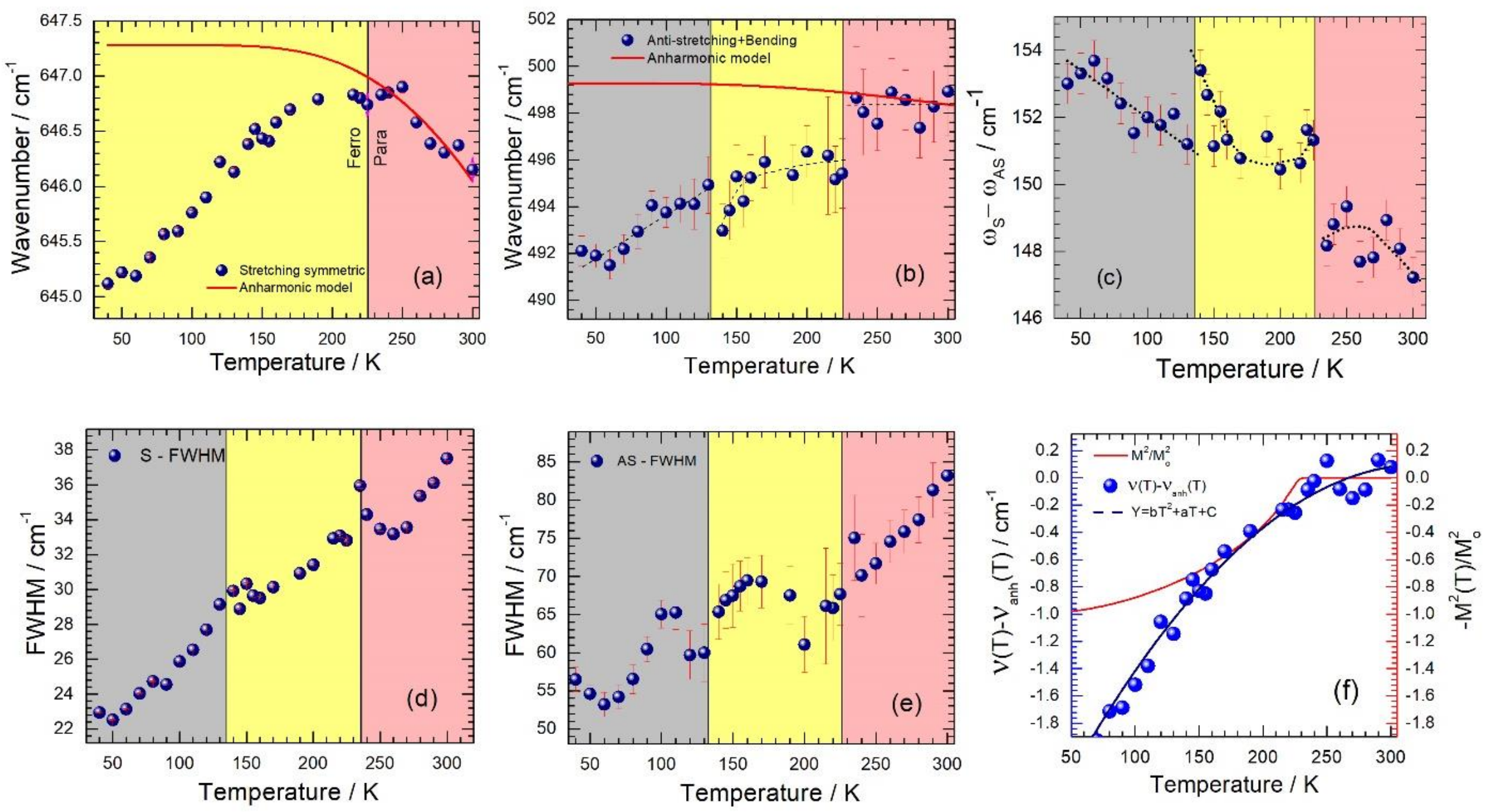

**Fig. 3** – (a-b) Temperature dependence of the symmetric (a) and antisymmetric stretching (b) wavenumbers observed for LCMO (blue spheres). The red lines show the usual behavior of the anharmonic effect contribution to the temperature dependence according to Balkanski's model. This line is not a fit in the temperature range in which the sample is ferromagnetic (in this region, it is an extrapolation following Balkanski's model). (c) Difference between AS and SS modes wavenumbers**.** Temperature dependence of full width at half maximum (FWHM) of the (d) SS (e) AS modes. (f) Temperature dependence of the departure from the anharmonic behavior of the mode that exhibits the spin-phonon coupling compared with $(M\,(T)/M_o)^2$.

The AS mode also exhibited an anomalous softening near the main magnetic transition (high-Tc), consistent with the reported spin-phonon coupling. However, we can also notice a small second discontinuity at about 135 K (see Fig. 3 (b)). In addition, the frequency difference between the SS and AS modes $\omega_{SS} - \omega_{AS}$ showed three distinct regions, with changes at about 226 and 135 K (see Fig. 3 (c)). As reported previously, this frequency difference is usually used as a comparative parameter of ordering [16].

The previous caracterization already clarified that the LCMO sample is partially disordered with $Co^{2+}$ and $Mn^{4+}$ cations. However, it is well known that, depending on structural order or oxygen content, LCMO can show two or more remarkable magnetic transitions due to: (I) a long-range FM superexchange $Mn^{4+}$-O-$Co^{2+}$, with high $T_C \sim 230$K; (II) a second FM interaction between the intermediate-spin $Co^{3+}(t^5e^1)$ and the Jahn-Teller $Mn^{3+}$ cations, created by the introduction of oxygen vacancies, with secondary low $T_{c2}$ [18,23,62]; (III) AFM interactions due to antisite disorder about 133 K [21,63] and (IV) a glassy behavior observed below at 80 K [21,63,64] owing to the coexistence of FM and AFM interactions. The bulk sintered in this paper showed only a strong high $T_C \sim 226$ K in the magnetization measurement, which evidences a predominance of a long-range ferromagnetic ordering. However, a careful analysis of the full width at half maximum (FWHM) of the stretching modes points out three different regions in FWHM with discontinuities near 135 K and 226 K, as observed in Fig. 3 (d) and (e). We can see those anomalies at the same temperatures in the linewidth of the SS and AS modes.

This result shows that, although the FC and ZFC magnetic curves do not evidence antisite disorder in LCMO, there is a new magnetic transition in LCMO, which can be characterized as an inhomogeneous magnetic phase [65]. This kind of change in phonon lifetimes was already observed as evidence of spin-phonon coupling [35,66]. This result is consistent with the existence of a majority ferromagnetic (FM) phase, basically an ordered region, and similar regions with disordered clusters, where the antiferromagnetic (AFM) interaction from $Co^{2+}$-O-$Co^{2+}$, $Mn^{4+}$-O-$Mn^{4+}$ and APB are predominate. This result shows the power of Raman spectroscopy to indicate magnetic inhomogeneity via spin-phonon coupling in these perovskites. Even not observing this effect in the magnetization curve, we could detect it in Raman phonon parameters.

Finally, it is expected that the phonon renormalization due to the spin-phonon coupling depart of the measured position with relation to the expected position due only to the anharmonic temperature dependence, which is proportional to $(M\,(T)/M_o)^2$, where $M(T)$ and $M_o$ are the magnetization at temperatures T and 0 K, respectively, as predicted by the mean-field theory [67]. In Fig. 3(f), we showed that this model disagrees with the experimental data for temperatures down to 140 K, where the second transition is observed, clearly showing we have a new renormalization of the frequency at this transition that comes from the spin-phonon coupling. This unconventional behavior exhibited by LCMO can be associated with a competition between FM and AFM magnetic states promoted by the antisite disorder. Different values and signals of magnetic couplings $J_{ij}$ can yield different contributions to the phonon renormalization induced by magnetic ordering. Interestingly, the SS mode renormalization in partially disordered LCMO was very similar to the one observed in $Y_2CoMnO_6$ (YCMO) [66], another multiferroic double perovskite whose stretching mode renormalization has been associated with the competition between FM and AFM magnetic states due to their frustrated Ising spin chain (E*-type) with ↑↑-↓↓spin pattern [68]. In both compounds (LCMO and YCMO), the different values of $J_{ij}$ caused an SS softening behavior that has a dependence on the square of the temperature, as observed in Fig. 3(f).

## 4. Conclusion

Summarizing, in this paper, we showed that Raman spectroscopy is an efficient tool to detect low disorder levels in superexchange driven ferromagnetic double perovskites. By monitoring the spin-phonon coupling in LCMO ceramic obtained by temperature-dependent Raman spectra, we showed an anomaly at 135 K due to antiferromagnetic (AF) interactions in LCMO, which can be probed even when the disorder levels are low. At the same time, in magnetization measurements, only a single magnetic transition could be observed. This result presents Raman spectroscopy as a powerful tool to probe disorder in double perovskites.

## Acknowledgments

The authors are grateful to CNPq (UNIVERSAL 431943/2016-8), CAPES, FAPEMIG, FUNCAP and FAPEMA (UNIVERSAL-01290/2016; INFRA-02050/21; BPD - 05073/21) for co-funding this work.

## References

[1] M.A. Magray, M. Ikram, Dielectric and Raman spectroscopy study of structural phase transformation of Sr-doped La2CoMnO6 double perovskite, J. Mater. Sci. Electron. 30 (2019) 8655–8666. doi:10.1007/s10854-019-01188-1.

[2] S. Das, R.C. Sahoo, S. Mishra, D. Bhattacharya, T.K. Nath, Effects of Ni doping at Co-site on dielectric, impedance spectroscopy and AC-conductivity in La2CoMnO6 double perovskites, Appl. Phys. A. 128 (2022) 354. doi:10.1007/s00339-022-05489-x.

[3] N.S. Rogado, J. Li, A.W. Sleight, M.A. Subramanian, Magnetocapacitance and Magnetoresistance Near Room Temperature in a Ferromagnetic Semiconductor: La2NiMnO6, Adv. Mater. 17 (2005) 2225–2227. doi:10.1002/adma.200500737.

[4] K.D. Chandrasekhar, A.K. Das, A. Venimadhav, Spin glass behaviour and extrinsic origin of magnetodielectric effect in non-multiferroic La2NiMnO6 nanoparticles., J. Phys. Condens. Matter. 24 (2012) 376003. doi:10.1088/0953-

8984/24/37/376003.

[5] P.R. Mandal, A. Khan, T.K. Nath, Antisite disorder driven magnetodielectric and magnetocaloric effect in double perovskite La2-xSrxCoMnO6 (x=0.0, 0.5, 1.0), J. Appl. Phys. 128 (2020). doi:10.1063/1.5135305 WE - Science Citation Index Expanded (SCI-EXPANDED).

[6] I. Álvarez-Serrano, M.L. López, F. Rubio, M. García-Hernández, G.J. Cuello, C. Pico, et al., Non-symmetric superparamagnetic clusters in the relaxor manganites Sr2−xBixMnTiO6 (0 ≤ x ≤ 0.75), J. Mater. Chem. 22 (2012) 11826. doi:10.1039/c2jm30670k.

[7] M.G. Masud, A. Ghosh, J. Sannigrahi, B.K. Chaudhuri, Observation of relaxor ferroelectricity and multiferroic behaviour in nanoparticles of the ferromagnetic semiconductor La(2)NiMnO(6)., J. Phys. Condens. Matter. 24 (2012) 295902. doi:10.1088/0953-8984/24/29/295902.

[8] Z. Meng, J. Xu, P. Yu, X. Hu, Y. Wu, Q. Zhang, et al., Double perovskite La2CoMnO6 hollow spheres prepared by template impregnation for high-performance supercapacitors, Chem. Eng. J. 400 (2020) 125966. doi:10.1016/j.cej.2020.125966.

[9] K. Yi, Q. Tang, Z. Wu, X. Zhu, Unraveling the Structural, Dielectric, Magnetic, and Optical Characteristics of Nanostructured La2NiMnO6 Double Perovskites, Nanomater. . 12 (2022). doi:10.3390/nano12060979.

[10] D. Yang, W. Wang, T. Yang, G.I. Lampronti, H. Ye, L. Wu, et al., Role of spontaneous strains on the biphasic nature of partial B-site disorder double perovskite La2NiMnO6, APL Mater. 6 (2018) 66102. doi:10.1063/1.5031486.

[11] J.E.F.S. Rodrigues, D. Morais Bezerra, A. Pereira Maciel, C.W. a. Paschoal, Synthesis and structural ordering of nano-sized Ba3B′Nb2O9 (B′ = Ca and Zn) powders, Ceram. Int. 40 (2014) 5921–5930. doi:10.1016/j.ceramint.2013.11.037.

[12] A. Dias, L.A. Khalam, M.T. Sebastian, C. William, C.W.A. Paschoal, R.L. Moreira, Chemical substitution in Ba(RE1/2Nb1/2)O-3 (RE = La, Nd, Sm, Gd, Tb, and Y) microwave ceramics and its influence on the crystal structure and phonon modes, Chem. Mater. 18 (2006) 214–220. doi:10.1021/cm051982f.

[13] A. Dias, R.L. Moreira, Far-infrared spectroscopy in ordered and disordered BaMg[sub 1/3]Nb[sub 2/3]O[sub 3] microwave ceramics, J. Appl. Phys. 94 (2003) 3414. doi:10.1063/1.1598634.

[14] J.E.F.S. Rodrigues, D.M. Bezerra, A.R. Paschoal, A.P. Maciel, C.W.D.A. Paschoal, Probing phase formation and structural ordering in Ba3ZnNb2O9 films using confocal Raman microscopy, Vib. Spectrosc. 72 (2014) 8–14. doi:10.1016/j.vibspec.2014.02.001.

[15] A.. Nowick, Y. Du, K.. Liang, Some factors that determine proton conductivity in nonstoichiometric complex perovskites, Solid State Ionics. 125 (1999) 303–311. doi:10.1016/S0167-2738(99)00189-7.

[16] R.X. Silva, A.S. Menezes, R.M. Almeida, R.L. Moreira, R. Paniago, X. Marti, et al., Structural order, magnetic and intrinsic dielectric properties of magnetoelectric

La2CoMnO6, J. Alloys Compd. 661 (2016) 541–552. doi:10.1016/j.jallcom.2015.11.097.

[17] P.A. Joy, Y.B. Khollam, S.K. Date, I. La, Spin states of Mn and Co in LaMn0.5Co0.5O3, Phys. Rev. B. 62 (2000) 8608–8610.

[18] R.I. Dass, J.B. Goodenough, Multiple magnetic phases of La2CoMnO6-δ (0~δ~0.05), Phys. Rev. B. 67 (2003) 014401. doi:10.1103/PhysRevB.67.014401.

[19] K.D. Truong, J. Laverdière, M.P. Singh, S. Jandl, P. Fournier, Impact of Co/Mn cation ordering on phonon anomalies in La2CoMnO6 double perovskites: Raman spectroscopy, Phys. Rev. B. 76 (2007) 132413. doi:10.1103/PhysRevB.76.132413.

[20] P.A. Joy, Y.B. Khollam, S.N. Patole, S.K. Date, Low-temperature synthesis of single phase LaMn0.5Co0.5O3, Mater. Lett. 46 (2000) 261–264.

[21] F. Liu, Y. Gao, H. Chang, Y. Liu, Y. Yun, Y. Gao, et al., Control of magnetic properties and band gap by Co / Mn ordering and, J. Magn. Magn. Mater. 435 (2017) 217–222. doi:10.1016/j.jmmm.2017.03.040.

[22] Y.Q. Lin, X.M. Chen, Dielectric, Ferromagnetic Characteristics, and Room-Temperature Magnetodielectric Effects in Double Perovskite La2CoMnO6 Ceramics, J. Am. Ceram. Soc. 94 (2011) 782–787. doi:10.1111/j.1551-2916.2010.04139.x.

[23] H.Z. Guo, A. Gupta, J. Zhang, M. Varela, S.J. Pennycook, Effect of oxygen concentration on the magnetic properties of La2CoMnO6 thin films, Appl. Phys. Lett. 91 (2007) 202509. doi:10.1063/1.2814919.

[24] R. Egoavil, S. Hühn, M. Jungbauer, N. Gauquelin, a Béché, G. Van Tendeloo, et al., Phase problem in the B-site ordering of La2CoMnO6: impact on structure and magnetism., Nanoscale. 7 (2015) 9835–43. doi:10.1039/c5nr01642h.

[25] C. Meyer, V. Roddatis, P. Ksoll, B. Damaschke, V. Moshnyaga, Structure, magnetism, and spin-phonon coupling in heteroepitaxial La2CoMnO6/Al2O3(0001) films, Phys. Rev. B. 98 (2018). doi:10.1103/PhysRevB.98.134433.

[26] A.J. Barón-González, C. Frontera, J.L. García-Muñoz, B. Rivas-Murias, J. Blasco, Effect of cation disorder on structural, magnetic and dielectric properties of La2MnCoO6 double perovskite., J. Phys. Condens. Matter. 23 (2011) 496003. doi:10.1088/0953-8984/23/49/496003.

[27] S. Yáñez-Vilar, M. Sánchez-Andújar, J. Rivas, M.A. Señarís-Rodríguez, Influence of the cationic ordering in the dielectric properties of the La2MnCoO6 perovskite, J. Alloys Compd. 485 (2009) 82–87. doi:10.1016/j.jallcom.2009.05.103.

[28] A.P. Ayala, I. Guedes, E.N. Silva, M.S. Augsburger, M. del C. Viola, J.C. Pedregosa, Raman investigation of A[sub 2]CoBO[sub 6] (A=Sr and Ca, B=Te and W) double perovskites, J. Appl. Phys. 101 (2007) 123511. doi:10.1063/1.2745088.

[29] M.C. Castro, E.F.V. Carvalho, W. Paraguassu, A.P. Ayala, F.C. Snyder, M.W. Lufaso, et al., Temperature-dependent Raman spectra of Ba2BiSbO6 ceramics,

J. Raman Spectrosc. 40 (2009) 1205–1210. doi:10.1002/jrs.2263.

[30] K.D. Truong, M.P. Singh, S. Jandl, P. Fournier, Investigation of phonon behavior in Pr(2)NiMnO(6) by micro-Raman spectroscopy., J. Phys. Condens. Matter. 23 (2011) 052202. doi:10.1088/0953-8984/23/5/052202.

[31] M. Viswanathan, P.S.A. Kumar, V.S. Bhadram, C. Narayana, A.K. Bera, S.M. Yusuf, Influence of lattice distortion on the Curie temperature and spin-phonon coupling in LaMn0.5Co0.5O3, J. Physics-Condensed Matter. 22 (2010) 1–8. doi:10.1088/0953-8984/22/34/346006.

[32] M.N. Iliev, H. Guo, A. Gupta, Raman spectroscopy evidence of strong spin-phonon coupling in epitaxial thin films of the double perovskite La[sub 2]NiMnO[sub 6], Appl. Phys. Lett. 90 (2007) 1–3. doi:10.1063/1.2721142.

[33] R.B. Macedo Filho, A. Pedro Ayala, C. William de Araujo Paschoal, Spin-phonon coupling in Y2NiMnO6 double perovskite probed by Raman spectroscopy, Appl. Phys. Lett. 102 (2013) 192902. doi:10.1063/1.4804988.

[34] K.D. Truong, M.P.P. Singh, S. Jandl, P. Fournier, Influence of Ni/Mn cation order on the spin-phonon coupling in multifunctional La2NiMnO6 epitaxial films by polarized Raman spectroscopy, Phys. Rev. B. 80 (2009) 134424. doi:10.1103/PhysRevB.80.134424.

[35] H.S. Nair, D. Swain, H. N., S. Adiga, C. Narayana, S. Elzabeth, Griffiths phase-like behavior and spin-phonon coupling in double perovskite Tb2NiMnO6, J. Appl. Phys. 110 (2011) 123919. doi:10.1063/1.3671674.

[36] R.X. Silva, H. Reichlova, X. Marti, D.A.B. Barbosa, M.W. Lufaso, B.S. Araujo, et al., Spin-phonon coupling in Gd(Co 1/2 Mn 1/2 )O 3 perovskite, J. Appl. Phys. 114 (2013) 194102. doi:10.1063/1.4829902.

[37] M.P. Pechini, Method of Preparing Lead and Alkaline Earth Titanates and Niobates and coating method using the same to form a capacitor. Patent Number: US3330697 A, Patent US 3330697, 1967.

[38] C.L. Bull, D. Gleeson, K.S. Knight, Determination of B-site ordering and structural transformations in the mixed transition metal perovskites La2CoMnO6 and La2NiMnO6, J. Phys. Condens. Matter. 15 (2003) 4927–4936. doi:https://doi.org/10.1088/0953-8984/15/29/304.

[39] A.C. Larson, R.B. Von Dreele, General Structure Analysis System (GSAS), Los Alamos Natl. Lab. Rep. LAUR. (2000) 86–748.

[40] B.H. Toby, EXPGUI , a graphical user interface for GSAS, J. Appl. Crystallogr. 34 (2001) 210–213. doi:10.1107/S0021889801002242.

[41] C.L. Bull, D. Gleeson, K.S. Knight, H. Search, C. Journals, A. Contact, et al., Determination of B-site ordering and structural transformations in the mixed transition metal perovskites La2CoMnO6 and La2NiMnO6, J. Phys. Condens. Matter. 15 (2003) 4927–4936. doi:10.1088/0953-8984/15/29/304.

[42] M.W. Lufaso, P.M. Woodward, Prediction of the crystal structures of perovskites using the software program SPuDS., Acta Crystallogr. B. 57 (2001) 725–38. doi:10.1107/S0108768101015282.

[43] X.L. Wang, M. James, J. Horvat, S.X. Dou, Spin glass behaviour in ferromagnetic La 2 CoMnO 6 perovskite manganite, Supercond. Sci. Technol. 15 (2002) 427–430. doi:10.1088/0953-2048/15/3/328.

[44] X.L.L. Wang, J. Horvat, H.K.K. Liu, A.H.H. Li, S.X.X. Dou, Spin glass state in Gd2CoMnO6 perovskite manganite, Solid State Commun. 118 (2001) 27–30. doi:10.1016/S0038-1098(01)00033-3.

[45] R. Dass, J. Goodenough, Multiple magnetic phases of La2CoMnO6-δ (0~δ~0.05), Phys. Rev. B. 67 (2003) 014401. doi:10.1103/PhysRevB.67.014401.

[46] R.P. Madhogaria, R. Das, E.M. Clements, V. Kalappattil, M.H. Phan, H. Srikanth, et al., Evidence of long-range ferromagnetic order and spin frustration effects in the double perovskite La2CoMnO6, Phys. Rev. B. 99 (2019) 1–13. doi:10.1103/PhysRevB.99.104436.

[47] M.A. Magray, M. Ikram, M. Najim, Impact of oxygen vacancies to control the magnetic and electronic properties of the La2CoMnO6 system, J. Magn. Magn. Mater. 529 (2021) 167857. doi:10.1016/j.jmmm.2021.167857.

[48] J. Blasco, J.L. García-Muñoz, J. García, G. Subías, J. Stankiewicz, J.A. Rodríguez-Velamazán, et al., Magnetic order and magnetoelectric properties of R2CoMnO6 perovskites ( R=Ho , Tm, Yb, and Lu), Phys. Rev. B. 96 (2017). doi:10.1103/PhysRevB.96.024409.

[49] J. Blasco, J. García, G. Subías, J. Stankiewicz, J.A. Rodríguez-Velamazán, C. Ritter, et al., Magnetoelectric and structural properties of Y2CoMnO6: The role of antisite defects, Phys. Rev. B. 93 (2016) 214401. doi:10.1103/PhysRevB.93.214401.

[50] J.Y. Moon, M.K. Kim, Y.J. Choi, N. Lee, Giant Anisotropic Magnetocaloric Effect in Double-perovskite Gd2CoMnO6 Single Crystals, Sci. Rep. 7 (2017) 1–10. doi:10.1038/s41598-017-16416-z.

[51] G. Blasse, Ferromagnetic interaction in non-metallic perovskites, J. Phys. Chem. Solids. 26 (1969) 1969–1971.

[52] J.F. Moulder, J. Chastain, Handbook of X-ray Photoelectron Spectroscopy: A Reference Book of Standard Spectra for Identification and Interpretation of XPS Data, Perkin Elmer Corp., 1992.

[53] H. Yang, J. Ouyang, A. Tang, Single step synthesis of high-purity CoO nanocrystals, J. Phys. Chem. B. 111 (2007) 8006–8013. doi:10.1021/jp070711k.

[54] J.K. Murthy, A. Venimadhav, Reentrant cluster glass behavior in La2CoMnO6 nanoparticles, J. Appl. Phys. 113 (2013) 163906. doi:10.1063/1.4802658.

[55] H. Guo, J. Burgess, S. Street, A. Gupta, T.G. Calvarese, M. a. Subramanian, Growth of epitaxial thin films of the ordered double perovskite La2NiMnO6 on different substrates, Appl. Phys. Lett. 89 (2006) 022509. doi:10.1063/1.2221894.

[56] A.T. Fulmer, J. Dondlinger, M.A. Langell, Passivation of the La2NiMnO6 double perovskite to hydroxylation by excess nickel , and the fate of the hydroxylated surface upon heating, Appl. Surf. Sci. 305 (2014) 544–553.

[57] M.N. Iliev, M. V Abrashev, A.P. Litvinchuk, V.G. Hadjiev, H. Guo, A. Gupta, Raman spectroscopy of ordered double perovskite La2CoMnO6 thin films, Phys. Rev. B. 75 (2007) 104118. doi:10.1103/PhysRevB.75.104118.

[58] M. Balkanski, R.F. Wallis, E. Haro, Anharmonic effects in light scattering due to optical phonons in silicon, Phys. Rev. B. 28 (1983) 1928–1934. doi:10.1103/PhysRevB.28.1928.

[59] D. Kumar, S. Kumar, V.G. Sathe, Spin-phonon coupling in ordered double perovskites A2CoMnO6 (A = La , Pr , Nd ) probed by micro-Raman spectroscopy., Solid State Commun. 194 (2014) 59–64. doi:10.1016/j.ssc.2014.06.017.

[60] M.N. Iliev, P. Padhan, A. Gupta, Temperature-dependent Raman study of multiferroic Bi2NiMnO6 thin films, Phys. Rev. B. 77 (2008) 172303. doi:10.1103/PhysRevB.77.172303.

[61] M.N. Iliev, M.M. Gospodinov, M.P. Singh, J. Meen, K.D. Truong, P. Fournier, et al., Growth, magnetic properties, and Raman scattering of La[sub 2]NiMnO[sub 6] single crystals, J. Appl. Phys. 106 (2009) 023515. doi:10.1063/1.3176945.

[62] M.P. Singh, K.D. Truong, P. Fournier, Magnetodielectric effect in double perovskite La2CoMnO6 thin films, Appl. Phys. Lett. 91 (2007) 042504. doi:10.1063/1.2762292.

[63] H. Chang, Y. Gao, F. Liu, Y. Liu, H. Zhu, Y. Yun, Effect of synthesis on structure, oxygen voids, valence bands, forbidden band gap and magnetic domain configuration of La2CoMnO6, J. Alloys Compd. 690 (2017) 8–14. doi:10.1016/j.jallcom.2016.08.086.

[64] H.Z. Guo, A. Gupta, T.G. Calvarese, M.A. Subramanian, Structural and magnetic properties of epitaxial thin films of the ordered double perovskite La2CoMnO6 Structural and magnetic properties of epitaxial thin films of the ordered, Appl. Phys. Lett. 89 (2006) 262503. doi:10.1063/1.2422878.

[65] I.O. Troyanchuk, A.P. Sazonov, H. Szymczak, Phase Separation in La 2-x Ax CoMnO6 (A = Ca and Sr) Perovskites, J. Exp. Theor. Phys. 99 (2004) 363–369.

[66] R.X. Silva, M.C. Castro Junior, S. Yánez-Vilar, M.S. Andújar, J. Mira, M.A. Señar?ís-Rodríguez, et al., Spin-phonon coupling in multiferroic Y2CoMnO6, J. Alloys Compd. 690 (2017) 909–915. doi:10.1016/j.jallcom.2016.07.010.

[67] E. Granado, A. García, J. Sanjurjo, C. Rettori, I. Torriani, F. Prado, et al., Magnetic ordering effects in the Raman spectra of La1-xMn1-xO3, Phys. Rev. B. 60 (1999) 11879–11882. doi:10.1103/PhysRevB.60.11879.

[68] T. Jia, Z. Zeng, X.G. Li, H.Q. Lin, The magnetic origin of multiferroic Y2CoMnO6, J. Appl. Phys. 117 (2015) 20–24. doi:10.1063/1.4915095.